\documentclass[twocolumn,showpacs,preprintnumbers,amsmath,amssymb]{revtex4}

\usepackage{graphicx}% Include figure files
\usepackage{dcolumn}% Align table columns on decimal point

\begin{document}

%\title{Biological-like molecular recognition effects at nanoscale}
%\title{Molecular recognition effects in the surface diffusion of large organic molecules}
%\title{Complex Organic Molecules and Cu Surface Interaction: A Theoretical and Experimental
%Study}
%\title{Non-biological Lock-and-Key effect on Cu surfaces: A Molecular Dynamics study}
%\title{Lock-and-Key like effect in non-biological systems: A Molecular Dynamics Study}
\title{Molecular Recognition Effects in the Surface Diffusion of Large Organic Molecules: The Case of Violet Lander }

\author{F. Sato$^{1}$}
\author{S. B. Legoas$^{2}$}
\author{R. Otero$^{3}$}
\author{F. H\"ummelink$^{3}$}
\author{P. Thostrup$^{3}$}
\author{E. L{\ae}gsgaard$^{3}$}
\author{I. Stensgaard$^{3}$}
\author{F. Besenbacher$^{3}$}
\author{D. S. Galv\~ao\footnote{Author to whom correspondence should be addressed. FAX:+55-19-35215376. Email: galvao@ifi.unicamp.br}$^{,1}$}

\affiliation{$^{1}$Instituto de F\'{\i}sica ``Gleb Wataghin", Universidade Estadual de Campinas, C.P. 6165, 13083-970 Campinas SP, Brazil}
\affiliation{$^{2}$Departamento de F\'{\i}sica, Universidade Federal de Roraima, 69304-000 Boa Vista RR, Brazil}
\affiliation{$^{3}$Interdisciplinary Nanoscience Center (iNANO) and Departament of Physics and Astronomy, University of Aarhus, DK-8000 Aarhus C, Denmark}

\date{\today}

\begin{abstract}
Violet Lander (VL) (C$_{108}$H$_{104}$) is a large organic molecule that when deposited on Cu (110) exhibited lock-and-key like behavior (Otero \textit{et al.}, Nature Mater. {\bf 3}, 779 (2004)). In this work we report on a detailed fully atomistic molecular dynamics study of this phenomenon. Our results show that it has its physical basis in the interplay of the molecular hydrogens and the Cu(110) atomic spacing, which is a direct consequence of an accidental commensurability between molecule and surface dimensions. This knowledge could be used to engineer new molecules capable of displaying lock-and-key behavior with new potential applications in nanotechology.
\end{abstract}

\pacs{68.65.-k, 61.46.+w, 68.37.Lp, 71.15.-m}
\clearpage
\maketitle

With the advent of nanoscience and nanotechnology and the perspective of molecular electronics \cite{molec,joach,opto,kuhn,oren,nanosen,mores}, significant theoretical and experimental efforts have been devoted to the study of the complex interactions involving organic molecular  structures and metallic surfaces \cite{yoko,rosei,rosei1,mores1,barth,theo,erem,otero}. One essential aspect of these phenomena is to understand how these interactions alter the properties of both molecule and surface. Recent progress has been achieved through the use of UHV-STM (ultrahigh vacuum-scanning transmission microscopy) \cite{rosei1,besen} that allowed the identification of important structural and dynamical features related to the behavior of molecular wires adsorved on metallic surfaces.

\begin{figure}[htbp]
\includegraphics[width=\columnwidth]{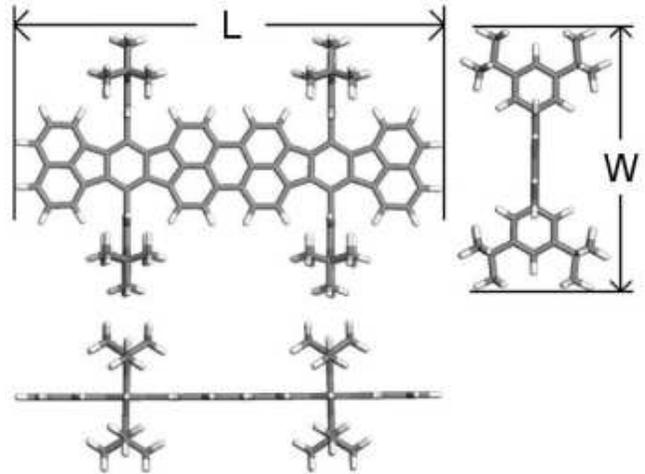}
\caption{Structural representation of the Violet Lander (C$_{108}$H$_{104}$) along different direction views. The experimental and theoretical values of the major molecular dimensions L and W are displayed in Table I (see text for discussions).}
\label{fig01}
\end{figure}

One important family of molecular wires is the so-called ``Lander molecules'' (because of its resemblance to a Mars surface rover) \cite{gho}. These large organic molecules are composed of a rigid polyaromatic $\pi$ central board and four spacers ({\it legs}) of up to eight 3,5-di-tert-butylphenyl $\sigma$-bonded to the central board (Fig.~\ref{fig01}). These spacers generate a configuration where the phenyl groups are nearly perpendicular to the main board plane by steric crowding. When deposited onto a metallic surface, these conformations allow an electronic decoupling ($\pi$-bonded from the metallic surface) to occur. Also, the presence of the tert-butyl groups, which increases the board-surface distance, permits some rotation of the ``legs'' without significantly reducing this distance.

\begin{figure}[hbtp]
\includegraphics[width=\columnwidth]{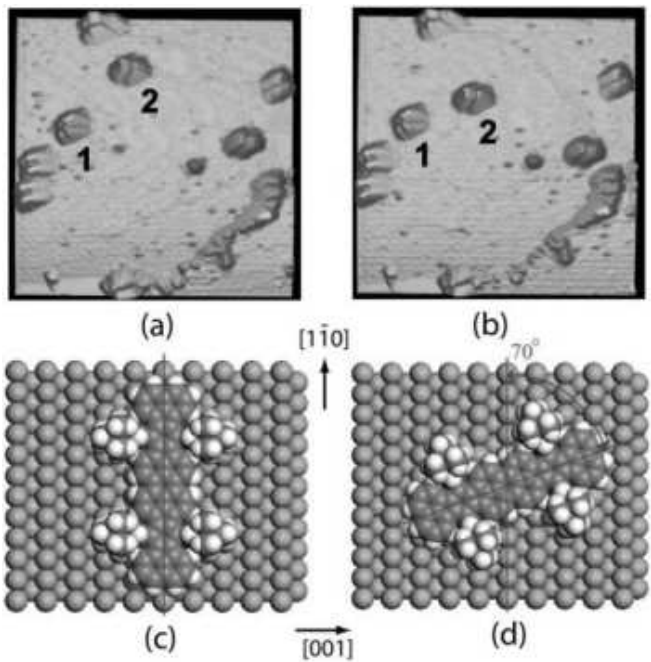}
\caption{(color online) STM snapshots from the scanning of VL molecules adsorbed on Cu(110) surface. (a)-(b) Molecules with their main boards aligned with the [1$\bar 1$0] direction (labeled 1) do not diffuse; molecules rotated by 70$^o$ (labeled 2) are able to diffuse along the [1$\bar 1$0] direction. (c)-(d) 3D-graphical atomistic representation of the VL in its (c) aligned and (d) 70$^o$ rotated configurations with respect to the [1$\bar 1$0] Cu(110) surface direction.}
\label{fig02}
\end{figure}

When adsorbed on Cu(100) or Cu(110) surfaces, the Lander molecule can act as a template for self-accomodating metal atoms at the step edges of the copper substrate, spontaneously generating metallic nanostructures dimensionally commensurable with the Lander \cite{otero,otero2}. This phenomenon opens up interesting possibilities for the bottom-up design of metallic nanostructures.

Similar studies for other molecules of the Lander family, the so-called Violet Lander (VL) (Violet due to its color) - C$_{108}$H$_{104}$ - led to the observation in the solid state of the first non-biological lock-and-key like effect \cite{lock}. A change by two orders of magnitude of the diffusion coefficients was observed when the molecular orientation of the substrate was changed. Previous experimental and theoretical investigations \cite{otero,lock,zambel} have revealed that VL molecules adsorb at room temperature (RT) on the Cu(110) surface with the polyaromatic board parallel to the substrate and aligned along the [1$\bar 1$0] direction (parallel configuration) [Fig.~\ref{fig02}(a), (c)]. In this configuration the molecules do not diffuse during the STM observation time (several minutes). The estimated diffusion coefficient is less than 5 $\times$ 10$^{-19}$cm$^2$s$^{-1}$  \cite{lock}.

When the STM tip is used as a tool to push the adsorbed molecule onto the copper surface at low temperatures (160-200 K) \cite{lock}, the molecule rotates 70$^o$ (from the [1$\bar 1$0] direction) [Fig.~\ref{fig02}(b), (d)], and the molecule diffuses along the [1$\bar 1$0] direction with an estimated diffusion coefficient of 4.8 $\times$ 10$^{-17}$ cm$^2$s$^{-1}$ \cite{lock}. This represents an increase of two orders of magnitude in relation to the parallel configuration. Interestingly, this molecular diffusion can be stopped at any time simply by flipping the molecule back to its parallel configuration. These two distinct configurations work as a two-state system (\texttt{0} and \texttt{1}  /  \textit{on} and \textit{off}), and it is the analogous of the so-called \textit{biological lock-and-key} recognition between enzymes and the substrate on which they act \cite{lock1}.

Surprisingly, in spite of many years of investigation into the diffusion of organic molecules on metallic surfaces, this phenomenon had not been observed before. As we shall see in the following discussions, its physical roots are based on an accidental commensurability of the VL and the atomic spacing of the Cu(110). Changing the molecule or/and the crystallographic direction, the phenomenon will not be observed.

In this work we present a detailed fully atomistic molecular dynamics study of the diffusion process of VL molecules adsorbed on a Cu(110) surface. We carried out molecular dynamics simulations in the framework of classical mechanics with standard force field \cite{cerius}, which includes van der Waals, bond stretch, bond angle bend, and torsional rotation terms. This methodology has been proven to be very effective for the study of dynamical properties of complex structures \cite{osc,c60,scro,gal}.

For all simulations the following convergence criteria were applied: maximum force of 0.005 kcal/mol/\AA, root mean square (RMS) deviations of 0.001 kcal/mol/\AA, energy differences of 0.0001 kcal/mol, maximum atomic displacement of 0.000 05 \AA, and RMS displacement of 0.000 01 \AA. A selective microcanonical (constant number of particles, volume, and total energy) impulse dynamics was used, with time steps of 1 fs.

Initially, the VL molecule was optimized in the gas phase (isolated). The use of full quantum methods for the whole system is not possible due to its large size, but the calculations for the isolated molecules is possible. In order to test the quality of the geometrical results for the molecular structures we have carried out a series of calculations using different methods: the semi-empirical hamiltonian AM1 available in the GAMESS package \cite{gamess}, and the \textit{ab initio} density functional methods Siesta \cite{siesta13,sanches-portal2004} and state-of-the-art DMOL3 \cite{dmol3-a,dmol3-b} in the local density approximation. The obtained structural dimensions are very consistent and in very good agreement with the estimated experimental data from STM experiments \cite{otero,lock,otero1,vl-exp-a}, indicating that the used molecular force field reproduces the VL geometrical features quite well.

%************************************************
\begin{table}
\begin{center}
\begin{tabular}{ccc}
\hline
\hline
Method & L (\AA) & W (\AA) \\
\hline
Molecular Mechanics & 25.34 & 15.45 \\
\hline
AM1 & 25.35 & 15.45 \\
\hline
Siesta & 25.25 & 15.48 \\
\hline
DMol3 & 25.16 & 15.36 \\
\hline
\hline
\end{tabular}
\caption{Violet Lander dimensions (Fig. 1), in {\AA}, optimized with classical molecular mechanics (univeral force field \cite{cerius}), semi-empirical AM1 method \cite{gamess}, DFT-LDA-Siesta \cite{siesta13,sanches-portal2004}, and DFT-LDA-Dmol3 \cite{dmol3-a, dmol3-b, dmol3-c}.} 
\end{center}
\end{table}
%************************************************

Then the molecule with its board parallel to the Cu(110) surface is placed (about $\sim$ 4-6 {\AA}) above the surface and set free to interact with the surface. We have also considered the cases where a little vertical impulse was given to the molecules to move them towards the surface. This is an additional test in order to provide enough kinetic energy to probe to local minima. These procedures were repeated varying the relative angular orientation and the geometries reoptimized. As the copper surfaces do not reconstruct, the Cu atoms were kept frozen at the experimental lattice value of $a = 3.61$ {\AA} \cite{cooper}.

\begin{figure}[hbtp]
\includegraphics[width=\columnwidth]{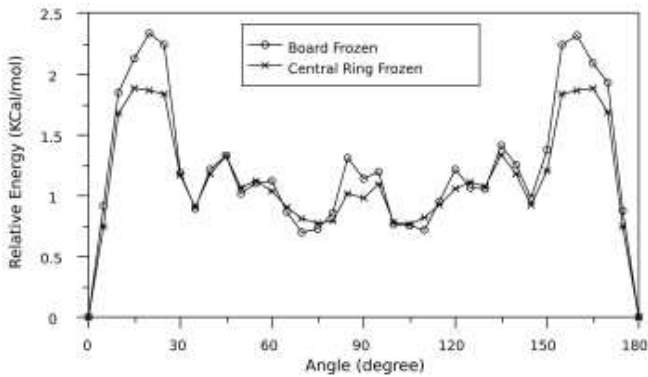}
\caption{Relative total energy profile of a VL molecule deposited on a Cu(110) surface as a function of the angle between the main axis of the molecule and the [1$\bar 1$0] direction (see Fig. 2). For each angle, the molecule is optimized with its  board or  central ring being frozen.}
\label{fig03}
\end{figure}

Our results showed that the molecules always converge to two possible configurations, 0 and 70$^o$ with respect to the  [1$\bar 1$0] orientation. In order to better understand the problem, we mapped the energy configuration as a function of the rotation angle. For comparative purposes this was carried out in two different ways: the molecule was placed at 0$^o$ and initially with the board frozen, then rotated in steps of 5$^o$ and the geometry optimized. The procedure was then repeated, keeping frozen only the central ring of the board. The results are displayed in Fig.~\ref{fig03}, and they are basically similar for the two procedures. We observed that in fact the two most stable configurations are at 0 and 70$^o$ (in agreement with the experimental data \cite{lock}), with a difference in energy of 0.75 kcal/mol. A third minimum was obtained for about 30$^o$ (but not experimentally observed), but considering its depth and width it could be overcome thermally.

We repeated the process (frozen central ring and steps of 5$^o$) in the [100] and [111] Cu surface directions. The results are displayed in Fig. 4. As we can see from Fig. 4, in contrast to the [110] surface (Fig. 3), the energy peaks and vallyes are smaller and smoother. No well defined constrained configuration was obtained. We observed in the simulations that these valleys and peaks could be easily thermically overcome. As discused in the following this energy could easily be overcome thermally for the [100] and [111] surfaces, which prevents the lock-key like phenomenon observed for the [110] surface.

\begin{figure}[htbp]
\begin{center}
\includegraphics[width=\columnwidth]{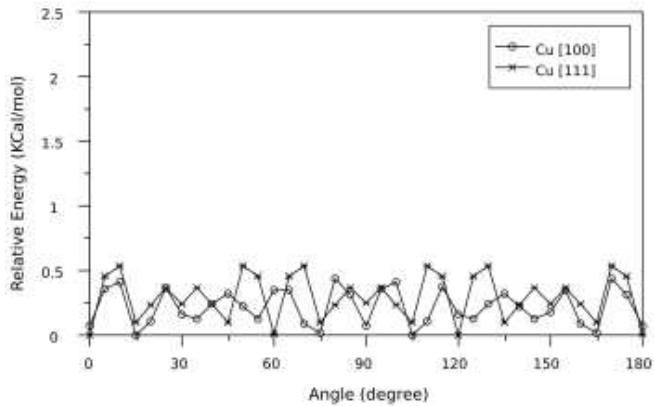}
\end{center}
\caption{Relative total energy profile of a VL molecule deposited on Cu[100] and Cu[111] surfaces as a function of the angle between the molecule main axis and the [010] and [$\bar 1$1$\bar 1$] direction, respectively.}
\label{fig04}
\end{figure}

The results show that the energy necessary for the Cu(110) surface diffusion in the non-rotated case is much higher than in the rotated case, and beyond the thermally available energy of the temperature at which the experiments are realized, while for the rotated case the energy necessary is thermically available.

Similarly to the rotational analysis (Fig. \ref{fig03}), we have also mapped the translational (keeping only the central board ring frozen) movement for the rotated and non-rotated configurations of the Cu[110] surface. Due to their energy profile mentioned above, the translational mappings do not provide new relevant information about the (100) and (111) surfaces

\begin{figure}[htbp]
\begin{center}
\includegraphics[width=\columnwidth]{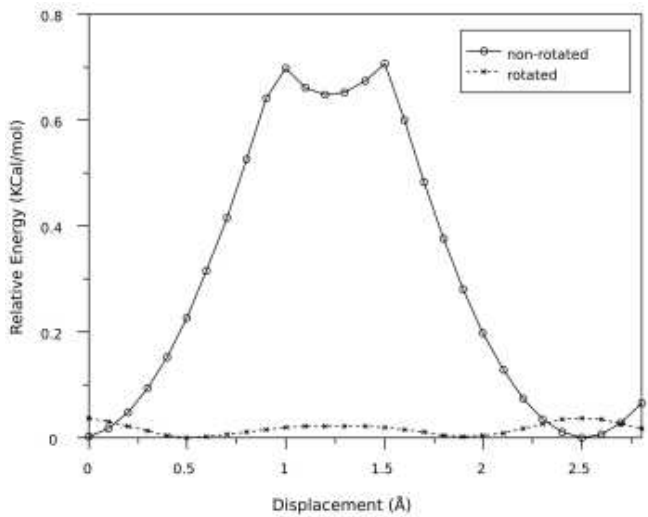}
\end{center}
\caption{Energy profile for the VL molecule displacement onto the Cu(110) surface, along the [1$\bar 1$0] direction in the rotated and non-rotated geometries.}
\label{fig05}
\end{figure}

The molecules were moved in steps of {0.1 \AA} along the  [1$\bar 1$0] direction, and for each point the geometry is optimized. The results are displayed in Fig. \ref{fig05}. Again, we observed that the energies associated with the movements of non-rotated and rotated cases are quite different, being higher for the non-rotated case. From Fig. 5 it is possible to explain why the molecules can easily diffuse in the rotated case. The reason for this being that we have a very low energy barrier for diffusion. However, although the barrier is much higher in the non-rotated case than in the rotated case, its absolute value is not high enough to prevent the overcoming of the potential energy barrier by kinetic energy. Thus, although the static analysis generated helpful information, it did not provide a complete physical description of the diffusion processes as the explicit inclusion of dynamical aspects is needed in order to explain the experimental data. Static procedure analysis has been often used in the literature. Our present analysis show that this procedure has limitations that could compromise the derived conclusions in some cases. These aspects have not been properly addressed before in the literature and might be of importance for the surface science of large organic adsorbates.

A deeper understanding of the molecular diffusion can be obtained from the analysis of the temporal evolution through  molecular dynamics simulations. Information on the force profile and molecular conformational changes in femto second scale can be easily addressed (not possible in the experimental case).

\begin{figure}[hbtp]
\begin{center}
\includegraphics[width=\columnwidth]{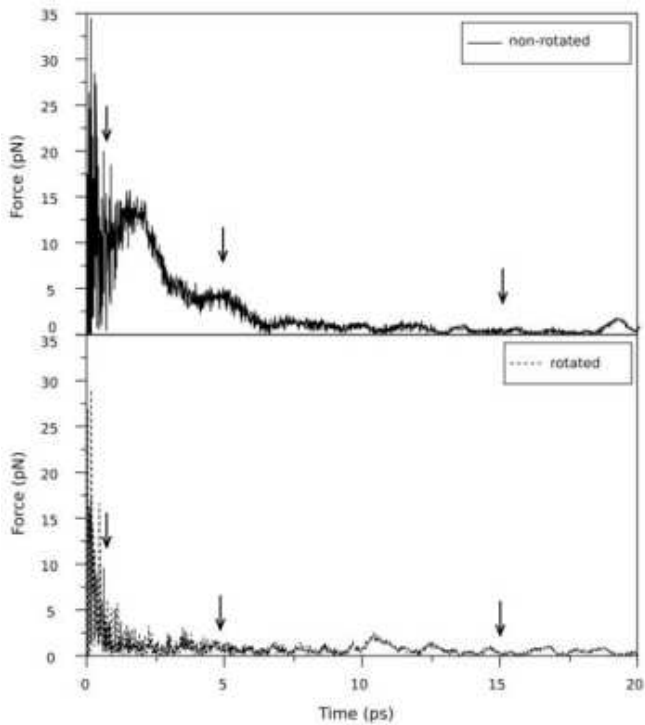}
\end{center}
\caption{Force profile, as a function of time, as a result of the interaction between the VL molecule in the (a) non-rotated and (b) rotated geometries, and the Cu(110) surface. Times corresponding to 1.8, 4.8, and 15 ps as a result of the initial impulse are shown by arrows.}
\label{fig06}
\end{figure}

We simulated the molecular diffusion of the VL molecules deposited onto a Cu(110) surface by impulse dynamics, i. e., we attributed an initial velocity to the molecule and followed its time evolution. This initial impulse is applied to mimic thermal effects or/and tip ``kick''. We have considered the cases of the non-rotated and rotated configurations. In each case different impulse velocities were used to determine the threshold values to induce molecular movements along the  [1$\bar 1$0] direction. For the non-rotated case we concluded that initial values larger than {0.9\AA}/ps (equivalent to a kinetic energy of 1.36 kcal/mol) are necessary to induce molecular diffusion. On the other hand, for the rotated case only values of {0.4\AA}/ps (equivalent to 0.27 kcal/mol) are required.

In Fig. \ref{fig06} we present the force profiles as a function of time of the forces experienced by the molecule for the different configurations. The displayed data are for a situation where the initial impulse velocities were set up to {0.8\AA}/ps along the [1$\bar 1$0] direction (see complementary material video01 \cite{movies}). For the non-rotated configuration (Fig. \ref{fig06}a), in the first $\sim$ 1.8 ps the force exerted on the molecule attains high values near to 20 pN in the opposite direction to the initial impulse. The force values decrease quickly ($\sim$ 5.0 pN at 5.0 ps and $<2.5$ pN at 15 ps) as the initial translational energy is converted into vibrational and torsional molecular movements. The molecule oscillates back and forth around its initial position, but no diffusion (net displacement) is observed (see video01). The situation is quite different for the rotated case (Fig. \ref{fig06}b). The initial forces are quickly attenuated, and the molecule diffuses easily (see video01). 

From the video01 we can clearly see that for the non-rotated case the molecule oscillates back and forth without diffusing. This oscillatory behavior can also be seen in the root mean displacement (RDM) data (Fig. 7). The associated RDM diffusion coefficients \cite{diffusion}  are 9.1 $\times$ 10$^{-6}$cm$^2$s$^{-1}$ and 5.6 $\times$ 10$^{-4}$cm$^2$s$^{-1}$ for the non-rotated and rotated cases, respectively. These diffusion coefficients are obtained from the analysis of the impulse molecular dynamics trajectory simulations where the initial impulse kinetic energy is quickly redistributed to the torsion/vibration/deformation-lengths molecular modes. In the experiments after the initial "tip kick" the non-rotated molecule undergoes a orientational transition to the rotated configuration, and due to the available thermal bath it has enough kinetic energy to diffuse. Although the absolute values of the theoretical and experimental diffusion coefficients cannot be directly compared, their differences can. They are in excellent agreement showing the same two orders of magnitude differences (9.1 $\times$ 10$^{-6}$cm$^2$s$^{-1}$ and 5.6 $\times$ 10$^{-4}$cm$^2$s$^{-1}$ versus 5.0 $\times$ 10$^{-19}$cm$^2$s$^{-1}$ and 4.8 $\times$ 10$^{-17}$cm$^2$s$^{-1}$ for the non-rotated and rotated case, theoretical and experimental values, respectively, i. e., a factor of 100 difference. 

\begin{figure}[hbtp]
\includegraphics[width=\columnwidth]{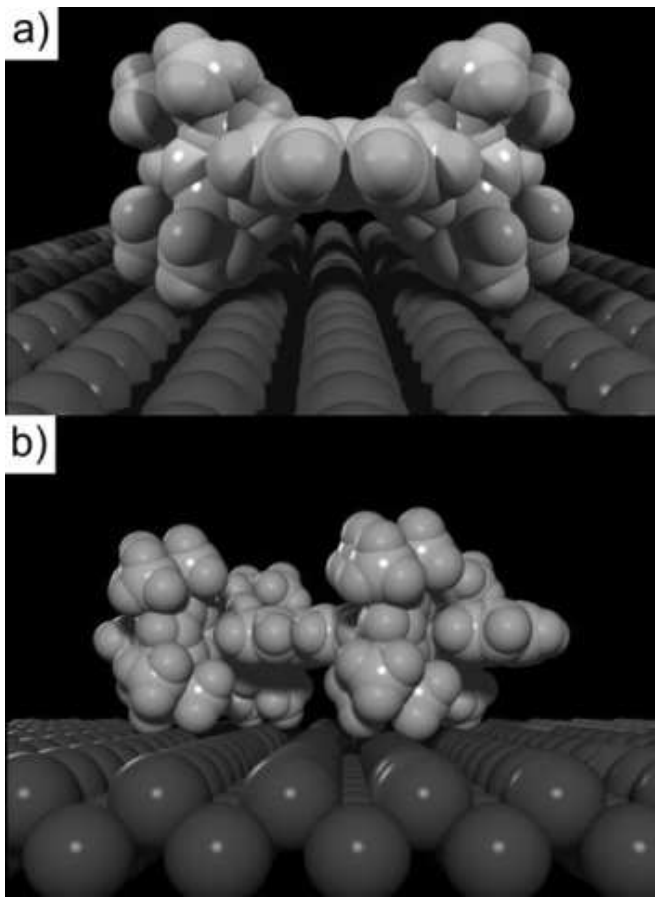}
\caption{Root mean displacement (RMD) of rotated (dot pointed curve) and non-rotated (fill curve) VL. The diffusion coefficient associated from curves are $9.1\times10^{-6} cm^{2}/s$ and $5.6\times10^{-4} cm^{2}/s$ for non-rotated and rotated VL, respectively.}
\label{fig07}
\end{figure}

\begin{figure}[hbt]
\includegraphics[width=7.0cm]{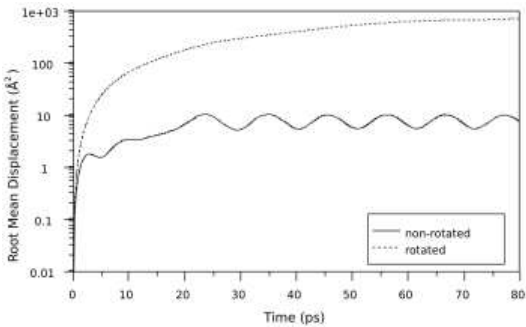}
\caption{(color online) Schematic view of a VL molecule in its (a) non-rotated, and (b) 70$^o$ rotated configurations with respect to the [1$\bar 1$0] Cu(110) surface direction.}
\label{fig08}
\end{figure}

\begin{figure}[hbt]
\includegraphics[width=7.0cm]{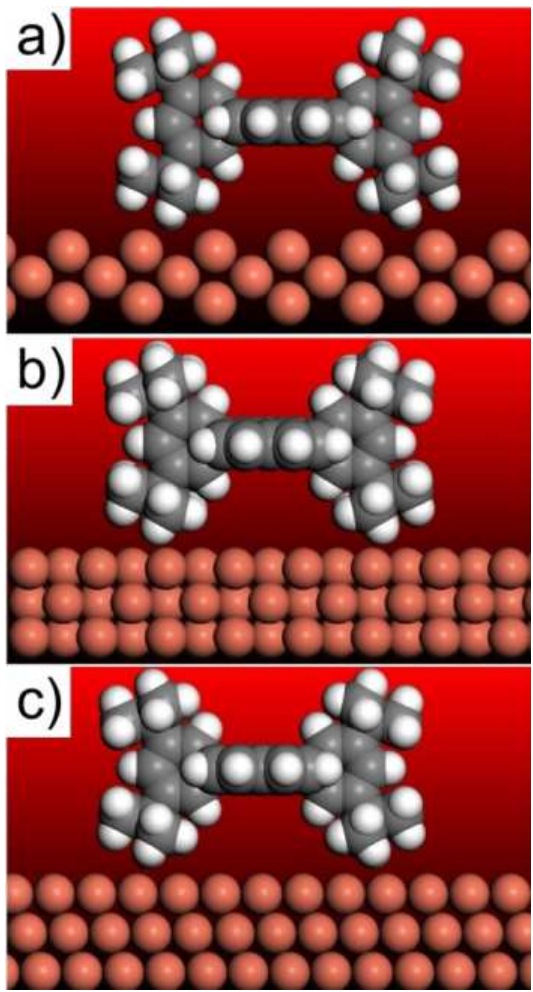}
\caption{(color online) Schematic view of the VL molecule for the rotated configuration of (a) [110]; (b) [100]; and (c) [111] directions. The different hydrogen atom orientations in relation to the Cu surface atoms are clearly visible.}
\label{fig09}
\end{figure}
 
The detailed analysis of the molecular conformational changes as a function of time allowed us to get a clear picture of the diffusional processes. The relative position of the hydrogen atoms of the molecular ``legs'' in relation to the copper (110) direction plays a fundamental role in defining whether diffusion is possible or not.

The VL adsorbed in the Cu(110) substrate exposes eight hydrogen atoms at the bottom of the legs. In the non-rotated configuration, these H atoms fit perfectly (``locked'') into the fourfold hollow sites of Cu(110) (Fig.~\ref{fig08}a). Due to this perfect fitting, any tentative translational energy is more likely to be converted into conformational changes than into net displacement (video01), thus blocking diffusion. For the rotated configuration, as the fitting is not as good (Fig.~\ref{fig08}b) as in Fig.~\ref{fig08}a, translational energy is easily converted into kinetic energy, and diffusion is possible.

It is important to stress that the configuration, in which diffusion is blocked, is just a consequence  of the accidental commensurability between the distance of the ``legs'' and the atomic spacing  of Cu(110). For instance, for the Cu(111) substrate the spacing is no longer  commensurate with the ``legs'' and the perfect ``fitting''  no longer occurs, and consequently the lock-and-key like effect ceases to exist (Fig. 9). The existence of the ``lucky'' commensurability in the case of VL/Cu(110) might explain why this effect was not observed erlier.

In principle, we can engineer other molecules to satisfy the commensurability criteria to obtain other  lock-and-key systems. As these molecules prove to be capable of exhibiting many interesting  properties (such as spontaneously building ordered nanostructures), this observation could be of great relevance to build nanostructures in a bottom-up approach. We hope the present study will stimulate further studies along these lines.

Work supported in part by the THEO-NANO, IN/MCT, IMMP/MCT, and Brazilian agencies CNPq, FINEP, and FAPESP.

\end{document}